\documentclass[aps,preprint]{revtex4}%
\usepackage{amssymb}
\usepackage{amsfonts}
\usepackage{amsmath}
\usepackage{graphicx}%
\providecommand{\U}[1]{\protect\rule{.1in}{.1in}}
\begin{document}
\title{Magnetic axis safety factor of finite $\beta$ spheromaks and transition from
spheromaks \ to toroidal magnetic bubbles}
\author{Paul M. Bellan}
\affiliation{Applied Physics and Materials Science, Caltech, Pasadena CA\ 91107, USA}
\author{Roberto Paccagnella}
\affiliation{Consorzio RFX and Istituto Gas Ionizzati del C.N.R, Corso Stati Uniti, 4 35127
Padova, Italy}
\keywords{one two three}\date{February 2,2015}

\begin{abstract}
The value of the safety factor on the magnetic axis of a finite-beta spheromak
is shown to be a function of beta in contrast to what was used in P. M.
Bellan, Phys. Plasmas \textbf{9}, 3050 (2002); this dependence on beta
substantially reduces the gradient of the safety factor compared to the
previous calculation. The method for generating finite-beta spheromak
equilibria is extended to generate equilibria describing toroidal magnetic
\textquotedblleft bubbles\textquotedblright\ where the hydrodynamic pressure
on the magnetic axis is \textit{less} than on the toroid surface. This
"anti-confinement" configuration can be considered an \ equilibrium with an
inverted beta profile and is relevant to interplanetary magnetic clouds as
these clouds have lower hydrodynamic pressure in their interior than on their surface.

\end{abstract}
\maketitle

\section{Introduction \ }

In Ref. \cite{Bellan2002}, one of the authors (PMB) examined analytic forms of
finite $\beta$ spheromak equilibria and used a well-known expression for the
value of the safety factor $q$ on the magnetic axis, denoted as $q_{axis}%
$,$\,\ $to argue that finite $\beta$ causes the beneficial effect of a much
larger $q\ $gradient than when $\beta=0$. However, co-author (RP) numerically
calculated $q_{axis}$ for these finite $\beta$ analytic equilibria and found
numerical results substantially different from the $q_{axis}$ given in
Ref.\cite{Bellan2002}. The reasons for this difference are identified as
resulting from a subtle misuse of an expression for $q_{axis}.$ Resolution of
this issue revealed that the analytic equilibria presented in
Ref.\cite{Bellan2002} could be extended to give an interesting toroidal
equilibria where $\ $ the pressure on the magnetic axis of a toroid is lower
than the pressure at the surface (edge) of the toroid rather than higher as in
a tokamak; i.e., the beta profile is inverted and the configuration is
bubble-like. Increase of a parameter $\bar{\gamma}$ (to be defined below)
results in solutions to a Grad-Shafranov equation evolving from characterizing
finite $\beta$ spheromak equilibria, to a conventional zero $\beta~$spheromak,
to magnetic \textquotedblleft bubbles\textquotedblright\ which are
tokamak-like configurations having inverted $\beta$ profiles, and then to a
tokamak with conventional $\beta$ profile. This evolution is characterized by
the ratio of two Bessel functions changing sign as their argument $\bar
{\gamma}$ is progressively increased. Interplanetary magnetic clouds are an
example of the magnetic bubble situation because on the magnetic axis these
clouds have lower hydrodynamic pressure than at their edge. Magnetic clouds
have been previously modeled using numerical solutions to Grad-Shafranov
equations \cite{Sonnerup2006},\cite{Hu2013} in a slab approximation (i.e.,
equations are solved in Cartesian geometry in the $xy$ plane with the $z$
direction ignorable); the model presented here differs by being analytic and
axisymmetric (i.e., equations are solved in cylindrical geometry in the $rz$
plane with the $\phi$ direction ignorable) so that, in contrast to a slab
approximation, toroidal geometry effects are inherently included. The analytic
model has only a few parameters and so has less freedom than a numerical model
but nevertheless has the useful feature of revealing parametric dependence and
scaling. The analytic model also offers the possibility of providing a useful
framework for other calculations, for example, calculating particle orbits in
an axisymmetric cloud; the virtues of developing a repertoire of analytic
solutions to the Grad-Shafranov equation has been discussed in
Ref.\cite{Cerfon2010}.

\section{Basic relations}

We use \ a cylindrical coordinate system $\{r,\phi,z\}$ and consider the
general axisymmetric magnetic field%
\begin{equation}
\mathbf{B}=\frac{1}{2\pi}\nabla\psi\times\nabla\phi+B_{\phi}r\nabla
\phi\label{B}%
\end{equation}
where $\psi$ is the poloidal flux function and $B_{\phi}$ is the toroidal
field. The $\phi$ direction is called the toroidal direction and any direction
lying in the poloidal plane ($rz$ plane) is called a poloidal direction. From
Ampere's law the associated current density is%
\begin{equation}
\mathbf{J}=\frac{1}{2\pi\mu_{0}}\nabla\left(  B_{\phi}r\right)  \times
\nabla\phi-\frac{r^{2}\nabla\phi}{2\pi\mu_{0}}\nabla\cdot\left(  \frac
{1}{r^{2}}\nabla\psi\right)  . \label{J}%
\end{equation}

We are interested in configurations where the poloidal flux function has a
local extremum in the $r,z$ plane; both spheromaks and tokamaks are this type
of configuration. The location of this extremum is called the magnetic axis
and its vertical location defines the $z$ origin while its radial location is
defined to be $r_{axis};$ $\psi$ is thus at a maximum or a minimum at
$r=r_{axis},$ $z=0.$ If $\psi$ is at a maximum on the magnetic axis then
$J_{\phi}$ is positive at the axis whereas if $\psi$ is at a minimum on the
magnetic axis then $J_{\phi}$ is negative at the axis.

Spheromaks and tokamaks are characterized by the safety factor $q$ which is
the number of times a field line goes around toroidally for each time it goes
poloidally around the magnetic axis. Tokamaks typically have near-unity $q$ on
the magnetic axis with $q$ increasing with increasing distance from the
magnetic axis whereas spheromaks have near-unity $q$ on the magnetic axis and
$q$ decreasing on moving away from the magnetic axis. The $\ $ gradient of
$q,$ denoted as $q^{\prime},$ provides stability properties and detailed
calculations show that a zero $\beta$ spheromak has small $q^{\prime}.$

The safety factor at the magnetic axis is given by \cite{Bellan-2000b}
\begin{equation}
q_{axis}=\ \frac{\ e^{1/2}+e^{-1/2}}{r_{axis}}\frac{B_{\phi,axis}}{\mu
_{0}J_{\phi,axis}\ }\ \label{qaxis}%
\end{equation}
where
\begin{equation}
e=\left(  \frac{\psi_{rr}}{\psi_{zz}}\right)  _{axis}\ \label{eps}%
\end{equation}
is a measure of the ellipticity of $\psi(r,z)~$in the vicinity of the magnetic
axis such that $e>1$ corresponds to vertically elongated equilibria (prolate)
while $e<1$ corresponds to vertically shortened equilibria (oblate). The
force-free relation $\mu_{0}J_{\phi,axis}/B_{\phi,axis}=\lambda$ was invoked
in Ref. \cite{Bellan-2000b} to give $q_{axis}=\left(  e^{1/2}+e^{-1/2}\right)
/(\lambda r_{axis})$ but this result is valid only if the plasma is indeed
force-free (i.e., has zero $\beta$ and equilibrium given by $\nabla
\times\mathbf{B}=\lambda\mathbf{B}$). If $\beta$ is finite, then\ $\mu
_{0}J_{\phi,axis}\neq\lambda B_{\phi,axis}$ and it is necessary to calculate
the actual value of $\mu_{0}J_{\phi,axis}/B_{\phi,axis}$ by consideration of
the details of the finite $\beta$ equilibrium.

To do this, we start by defining $\beta_{rel}$
\begin{equation}
\beta_{rel}=\mu_{0}\frac{P_{axis}-P_{lc}}{B_{axis}^{2}} \label{beta}%
\end{equation}
where $P_{axis}$ and $P_{lc}$ are respectively the hydrodynamic pressures on
the magnetic axis and on the last closed flux surface. Positive $\beta_{rel}$
thus corresponds to a conventional $\beta$ profile whereas negative
$\beta_{rel}$ corresponds to an inverted $\beta$ profile. This definition
differs from that used in Ref. \cite{Bellan2002} because (i)\ here
$B_{axis}^{2}$ is used and (ii)\textit{\ }a relative rather than absolute
pressure is used. \textit{ } The definition in Ref. \cite{Bellan2002} used, in
contrast, the \textit{average} \textit{poloidal} field linking the circular
surface lying in the $z=0$ plane between the geometric axis and the magnetic
axis. Because the definition of $\beta_{rel}$ uses the relative hydrodynamic
pressure$,$ it is seen that $\beta_{rel}$ can be positive or negative. In
particular, if $P_{axis}$ is smaller then $P_{lc},$ then $\beta_{rel}$ will be
negative$.$ The definition of $\beta_{rel}$ is useful because it provides a
simple mathematical way to distinguish toroidal equilibria with inverted
$\beta$ profiles $\ $from those with normal $\beta$ profiles.$\ $ The former
are toroidal magnetic bubbles while the latter are toroidal confinement
configurations such as spheromaks and tokamaks.

On expressing the magnetic field as
\begin{equation}
\mathbf{B}=\frac{1}{2\pi}\left(  \nabla\psi(r,z)\times\nabla\phi+\mu
_{0}I(r,z)\nabla\phi\right)  \label{B with current}%
\end{equation}
where $I=2\pi rB_{\phi}/\mu_{0}$ is the poloidal current, MHD equilibrium
$\mathbf{J\times B}=\nabla P$ can be expressed as the Grad-Shafranov equation
\cite{Grad1958,Shafranov1966}
\begin{equation}
r\frac{\partial}{\partial r}\left(  \frac{1}{r}\frac{\partial\psi}{\partial
r}\right)  +\frac{\partial^{2}\psi}{\partial z^{2}}+4\pi^{2}\mu_{0}r^{2}%
\frac{dP}{d\psi}+\mu_{0}^{2}I\frac{dI}{d\psi}=0. \label{GS}%
\end{equation}

We assume that $P$ is a linear function of the poloidal flux $\psi$ and so can
be expressed as
\begin{equation}
P=\frac{P_{axis}-P_{lc}}{\left(  \psi_{axis}-\psi_{lc}\right)  }\psi
-\frac{P_{axis}\psi_{lc}-P_{lc}\psi_{axis}}{\left(  \psi_{axis}-\psi
_{lc}\right)  } \label{P}%
\end{equation}
where $\psi_{lc}$ is the last closed flux surface of the configuration.

The poloidal current is similarly assumed to be a linear function of the
poloidal flux and can be expressed as
\begin{equation}
\mu_{0}I=\lambda\psi. \label{I}%
\end{equation}
We note that the assumed linear dependence in Eq.\ref{I} differs from the
assumption used in Solov'ev-type solutions such as in Ref.\cite{Cerfon2010}
\ where it is assumed that $I^{2}\sim\psi+const.$ For the linear dependence
assumed here, $IdI/d\psi$ is linear in $\psi$ whereas for the Solov'ev-type
assumption, $IdI/d\psi$ is a constant.

Using Eq.\ref{I}, the toroidal component of Eq.\ref{B with current} gives
\begin{equation}
B_{\phi,axis}=\frac{\lambda\psi_{axis}}{2\pi r_{axis}}. \label{Bphi axis}%
\end{equation}
The gradient of $P$ with respect to $\psi$ can then be expressed in terms of
$\beta_{rel}$ as%
\begin{equation}
\frac{dP}{d\psi}=\beta_{rel}\frac{B_{axis}^{2}}{\mu_{0}\psi_{axis}\left(
1-\psi_{lc}/\psi_{axis}\right)  }. \label{grad P}%
\end{equation}
Since $B_{pol}$ vanishes at the magnetic axis, $B_{axis}^{2}=B_{\phi,axis}%
^{2}$ and so
\begin{equation}
\frac{dP}{d\psi}=\beta_{rel}\frac{\lambda^{2}\psi_{axis}}{4\pi^{2}r_{axis}%
^{2}\mu_{0}\left(  1-\psi_{lc}/\psi_{axis}\right)  }. \label{grad P 2}%
\end{equation}
$\ $

\section{Cylindrical Solutions to Finite $\beta_{rel}$ Grad-Shafranov
Equation}

We now introduce dimensionless quantities%
\begin{equation}
\bar{\psi}=\frac{\psi}{\psi_{axis}},\text{ }\bar{r}=\frac{r}{r_{axis}},\text{
}\bar{z}=\frac{z}{r_{axis}},\text{ }\bar{\lambda}=\lambda r_{axis}
\label{norm}%
\end{equation}
so Eq.\ref{GS} can be expressed as%

\begin{equation}
\ \bar{r}\frac{\partial}{\partial\bar{r}}\left(  \frac{1}{\bar{r}}%
\frac{\partial\bar{\psi}}{\partial\bar{r}}\right)  +\ \frac{\partial^{2}%
\bar{\psi}}{\partial\bar{z}^{2}}+\ \bar{\lambda}^{2}\left(  \frac{\beta
_{rel}\bar{r}^{2}}{\ 1-\bar{\psi}_{lc}}+\ \bar{\psi}\right)  =0.
\label{GS norm}%
\end{equation}
We define
\begin{equation}
\bar{\chi}=\frac{\beta_{rel}\bar{r}^{2}}{\left(  1-\bar{\psi}_{lc}\right)
}+\bar{\psi} \label{chi}%
\end{equation}
so Eq.\ref{GS norm} becomes%
\begin{equation}
\ \bar{r}\frac{\partial}{\partial\bar{r}}\left(  \frac{1}{\bar{r}}%
\frac{\partial\bar{\chi}}{\partial\bar{r}}\right)  +\ \frac{\partial^{2}%
\bar{\chi}}{\partial\bar{z}^{2}}+\ \bar{\lambda}^{2}\bar{\chi}=0.
\label{chi GS}%
\end{equation}

We assume a solution of the form%
\begin{equation}
\bar{\chi}=\bar{r}g(\bar{r})\cos(\bar{k}\bar{z}) \label{def g}%
\end{equation}
so Eq.\ref{chi GS} becomes%

\begin{equation}
\ \frac{\partial^{2}g}{\partial\bar{r}^{2}}+\frac{1}{\bar{r}}\frac{\partial
g}{\partial\bar{r}}+\left(  \bar{\gamma}^{2}-\frac{1}{\bar{r}^{2}}\right)  g=0
\label{g eq}%
\end{equation}
where
\begin{equation}
\bar{\gamma}^{2}=\bar{\lambda}^{2}-\bar{k}^{2}. \label{gamma}%
\end{equation}
Equation \ref{g eq} is Bessel's equation with general solution for real
$\bar{\gamma}$
\begin{equation}
g(\bar{r})=\sigma_{J}J_{1}(\bar{\gamma}\bar{r})+\sigma_{Y}Y_{1}(\bar{\gamma
}\bar{r}) \label{solve g}%
\end{equation}
where $\sigma_{J}$ and $\sigma_{Y}$ are constant coefficients to be determined
by boundary conditions.

From Eqs.\ref{chi} and \ref{def g} the solution to the normalized
Grad-Shafranov equation is
\begin{equation}
\bar{\psi}=\bar{r}\left(  \sigma_{J}J_{1}(\bar{\gamma}\bar{r})+\sigma_{Y}%
Y_{1}(\bar{\gamma}\bar{r})\right)  \cos(\bar{k}\bar{z})-\bar{r}^{2}%
Q\ \label{GS soln}%
\end{equation}
where
\begin{equation}
Q=\frac{\beta_{rel}}{1-\bar{\psi}_{lc}}. \label{def Q}%
\end{equation}

\bigskip However, $\bar{\psi}=1$ is required at $\bar{r}=1,$ $\bar{z}=0$
(i.e., at the magnetic axis)\ so%
\begin{equation}
\sigma_{J}J_{1}(\bar{\gamma})+\sigma_{Y}Y_{1}(\bar{\gamma}%
)=1+Q\ .\ \label{intermediate 0}%
\end{equation}
The following three Bessel identities where $C_{n}=J_{n}$ or $Y_{n}$ will now
be used repeatedly in the rest of the discussion:
\begin{subequations}
\label{identities}%
\begin{align}
\frac{dC_{0}(s)}{ds}  &  =-C_{1}(s)\label{ident 1}\\
\frac{d}{ds}\left(  sC_{1}(s)\right)   &  =sC_{0}(s)\label{ident 2}\\
sC_{2}(s)\  &  =2C_{1}(s)-sC_{0}(s)\ . \label{identity}%
\end{align}
The magnetic axis is also where $\partial\bar{\psi}/\partial\bar{r}$ vanishes
and so taking the derivative of Eq.\ref{GS soln} with respect to $\bar{r},$
using Eq.\ref{ident 2}, and then setting $\bar{r}=1$ and $\bar{z}=0$ gives
\end{subequations}
\begin{equation}
\ \sigma_{J}J_{0}(\bar{\gamma})+\sigma_{Y}Y_{0}(\bar{\gamma})=\frac{2}%
{\bar{\gamma}}Q. \label{deriv GS}%
\end{equation}
Equations \ref{intermediate 0} and \ref{deriv GS} constitute two linear
inhomogeneous algebraic equations for the coefficients $\sigma_{J}$ and
$\sigma_{Y}.$ Solving these equations for $\sigma_{J}$ and $\sigma_{Y}$ and
using the Wronskian%
\begin{equation}
J_{1}(s)Y_{0}(s)-J_{0}(s)\ Y_{1}(s)\ =\frac{2}{\pi s} \label{Wronskian}%
\end{equation}
and Eq.\ref{identity} gives
\begin{subequations}
\label{sigmas}%
\begin{align}
\sigma_{J}  &  =\frac{\pi\bar{\gamma}}{2}\left(  Y_{0}(\bar{\gamma}%
)-QY_{2}\left(  \bar{\gamma}\right)  \right) \label{solve sigmaJ}\\
\sigma_{Y}  &  =\frac{\pi\bar{\gamma}}{2}\left(  -J_{0}(\bar{\gamma}%
)+QJ_{2}(\bar{\gamma})\right)  . \label{solve sigmaY}%
\end{align}

\section{Spheromak-type solutions}

Spheromaks are \textit{singly-connected} Grad-Shafranov equilibria (i.e.,
there is no \textquotedblleft hole\textquotedblright\ in the \textquotedblleft
doughnut\textquotedblright) and so the domain includes $\bar{r}=0.$ A
spheromak therefore cannot contain a $Y_{1}(\bar{\gamma}\bar{r})$ component
because $Y_{1}(\bar{\gamma}\bar{r})$ diverges at $\bar{r}=0.$ It is thus
necessary to impose $\sigma_{Y}=0$ for a spheromak in which case
Eq.\ref{solve sigmaY} yields the relation%
\end{subequations}
\begin{equation}
Q=\frac{J_{0}(\bar{\gamma})}{J_{2}(\bar{\gamma})}. \label{Q}%
\end{equation}
Substituting for $Q$ in Eq.\ref{solve sigmaJ} and using Eqs.\ref{identity} and
\ref{Wronskian} gives%
\begin{equation}
\sigma_{J}=\frac{2}{\bar{\gamma}J_{2}(\bar{\gamma})}. \label{sigmaJ spheromak}%
\end{equation}

Using Eq.\ref{identity} to substitute for $J_{2}(\bar{\gamma})$ in Eq.\ref{Q}
shows that Eq.\ref{Q} can alternately be written as%
\begin{equation}
Q=\frac{2J_{1}(\bar{\gamma})}{\bar{\gamma}J_{2}(\bar{\gamma})}-1
\label{solve Q}%
\end{equation}
so one can also write $\sigma_{J}$ as
\begin{equation}
\sigma_{J}=\ \frac{1+Q}{J_{1}(\bar{\gamma})}. \label{alt sigmaJ spheromak}%
\end{equation}
Because $\sigma_{Y}=0$ for a spheromak Eqs.\ref{def Q} and \ref{Q} show
$\ $that a spheromak has
\begin{equation}
\beta_{rel}=\frac{J_{0}(\bar{\gamma})}{J_{2}(\bar{\gamma})}\left(  1-\bar
{\psi}_{lc}\right)  \ \label{beta spheromak}%
\end{equation}
and
\begin{equation}
\sigma_{J}=\frac{1}{J_{1}(\bar{\gamma})}\left(  1+\frac{\beta_{rel}}%
{1-\bar{\psi}_{lc}}\right)  . \label{solve sigma}%
\end{equation}
On substituting for $\sigma_{J}$ and $Q$ in Eq.\ref{GS soln} the solution to
the normalized Grad-Shafranov equation becomes%
\begin{equation}
\bar{\psi}=\bar{r}\frac{J_{1}(\bar{\gamma}\bar{r})}{J_{1}(\bar{\gamma}%
)}\left(  1+\frac{\beta_{rel}}{1-\bar{\psi}_{lc}}\right)  \cos(\bar{k}\bar
{z})-\bar{r}^{2}\frac{\beta_{rel}}{1-\bar{\psi}_{lc}} \label{GS solve}%
\end{equation}

If $\beta_{rel}=0$ and $\bar{\psi}_{lc}=0\ $are additionally assumed, the
standard result for a zero-beta spheromak in a cylindrical flux conserver of
radius $a$ is retrieved, namely $\bar{\gamma}=x_{01}=2.405$ where $x_{01}$ is
the first root of $J_{0}.$ Since $\bar{\gamma}=\gamma r_{axis},~$and the last
closed flux surface is at the cylinder radius, then the assumption $\bar{\psi
}_{lc}=0$ and $\beta_{rel}=0$ in Eq. \ref{GS solve} implies $J_{1}(\gamma
a)=0$ in which case \ $\gamma a=x_{11}=3.83$ where $x_{11}$ is the first root
of $J_{1}.$ Thus, for a $\beta_{rel}=0$ spheromak, $r_{axis}/a=\bar{\gamma
}/(\gamma a)=x_{01}/x_{11}=\allowbreak0.63\ $as is well known. Equation
\ref{beta spheromak} shows that spheromaks with finite positive $\beta_{rel}$
are restricted to the range $0<\bar{\gamma}<2.405$ but, as will be discussed
in Sec.\ref{bubble}, physically relevant non-spheromak configurations with
negative $\beta_{rel}$ exist when $\bar{\gamma}>2.405.$

Substitution of Eq.\ref{beta spheromak} into Eq.\ref{GS solve} gives%
\begin{equation}
\bar{\psi}=\frac{1}{J_{2}(\bar{\gamma})}\left(  \frac{2\bar{r}}{\bar{\gamma}%
}J_{1}(\bar{\gamma}\bar{r})\cos(\bar{k}\bar{z})-\bar{r}^{2}J_{0}(\bar{\gamma
})\right)  \label{solve GS alt}%
\end{equation}
which reverts to the $\beta_{rel}=0$ solution when $\bar{\gamma}=2.405$ as can
be seen using Eq.\ref{identity} to give \linebreak$\bar{\gamma}J_{2}%
(\bar{\gamma})=2J_{1}(\bar{\gamma})$ if $J_{0}(\bar{\gamma})=0$.

\section{Safety Factor of Spheromaks with finite $\beta_{rel}$}

The last closed flux surface of a spheromak has $\bar{\psi}_{lc}=0$ and
$P_{lc}=0$ in which case Eqs. \ref{beta} and \ref{Bphi axis} give%
\begin{equation}
\beta_{rel}=4\pi^{2}r_{axis}^{2}\frac{\mu_{0}P_{axis}}{\lambda^{2}\psi
_{axis}^{2}} \label{beta spheromak 2}%
\end{equation}
and Eq.\ref{GS solve} becomes%
\begin{equation}
\bar{\psi}=\bar{r}\frac{J_{1}(\bar{\gamma}\bar{r})}{J_{1}(\bar{\gamma}%
)}\left(  1+\beta_{rel}\right)  \cos(\bar{k}\bar{z})-\beta_{rel}\bar{r}^{2}
\label{GS spheromak}%
\end{equation}
which is the same as Eq.(2) of Ref. \cite{Bellan2002} except for the different
definition of $\beta_{rel}.$

In order to determine $q_{axis},$ Eq.\ref{qaxis} \ shows that it is necessary
to calculate $\mu_{0}J_{\phi,axis}/B_{\phi,axis}.$ Equation \ref{J} shows that%
\begin{equation}
\mu_{0}J_{\phi}=-\frac{\psi_{axis}}{2\pi\bar{r}r_{axis}^{3}}\left[  \bar
{r}\frac{\partial}{\partial\bar{r}}\left(  \frac{1}{\bar{r}}\frac{\partial
\bar{\psi}}{\partial\bar{r}}\right)  +\ \frac{\partial^{2}\bar{\psi}}%
{\partial\bar{z}^{2}}\right]  \label{Jphi}%
\end{equation}
so, using Eq.\ref{Bphi axis} and Eq.\ref{GS norm} it is seen that%
\begin{equation}
\frac{\mu_{0}J_{\phi,axis}}{B_{\phi,axis}}=\left(  1+\beta_{rel}\right)
\lambda. \label{Jphi by Bphi}%
\end{equation}
Thus $\mu_{0}J_{\phi,axis}/B_{\phi,axis}=\lambda$ only if $\beta_{rel}=0.$
Inserting Eq.\ref{Jphi by Bphi} in Eq.\ref{qaxis} gives%
\begin{equation}
q_{axis}=\ \frac{\ e^{1/2}+e^{-1/2}}{\bar{\lambda}\left(  1+\beta
_{rel}\right)  }\ \label{solve q axis}%
\end{equation}
which differs from Eq.(30)\ of Ref.\cite{Bellan2002} by having an extra and
important factor of $(1+\beta_{rel})$ in the denominator.

From Eq.\ref{solve GS alt} and use of the Bessel identities it is seen that
\begin{subequations}
\label{psizzrr}%
\begin{align}
\left(  \bar{\psi}_{zz}\right)  _{axis}  &  =-\frac{2\bar{k}^{2}}{\bar{\gamma
}\ }\frac{J_{1}(\bar{\gamma})}{J_{2}(\bar{\gamma})}\label{psizz}\\
\left(  \bar{\psi}_{rr}\right)  _{axis}  &  =\frac{-2\bar{\gamma}J_{1}%
(\bar{\gamma})}{J_{2}(\bar{\gamma})} \label{psirr}%
\end{align}
so the ellipticity is%
\end{subequations}
\begin{equation}
e=\frac{\bar{\gamma}^{2}}{\bar{k}^{2}}. \label{solve eps}%
\end{equation}
This indicates that the poloidal flux surfaces will be circular near the
magnetic axis (i.e., have $e=1$) if $\bar{\gamma}=\bar{k}$ in which case
$\bar{\lambda}=\sqrt{2}\bar{\gamma}.$ Combination of Eqs.\ref{beta spheromak},
\ref{solve q axis}, and \ref{solve eps} gives
\begin{equation}
q_{axis}=\ \frac{\bar{\lambda}}{\bar{\gamma}\bar{k}}\frac{1}{1+\beta_{rel}%
}=\ \frac{\bar{\lambda}}{2\bar{k}}\frac{J_{2}(\bar{\gamma})}{J_{1}(\bar
{\gamma})}. \label{solve qaxis}%
\end{equation}
Equation \ref{solve qaxis} has been validated by direct numerical integration
of field lines in the vicinity of the magnetic axis of a magnetic
configuration characterized by Eq.\ref{B with current} with $\bar{\psi}$ given
by Eq.\ref{solve GS alt}. $\ $In the $\beta_{rel}=0\ $limit, $J_{0}%
(\bar{\gamma})=0$ and $q_{axis}\rightarrow\bar{\lambda}/\left(  \bar{\gamma
}\bar{k}\right)  $ which is Eq.(33) of Ref.\cite{Bellan2002}, but for finite
positive $\beta_{rel},$ Eq. \ref{solve qaxis} shows that $q_{axis}$ is reduced
from its $\beta_{rel}=0$ value.

The safety factor at the wall is \cite{Bellan2002}
\begin{equation}
q_{wall}=\frac{\bar{\lambda}}{2\pi\bar{k}}\cos^{-1}\left(  J_{0}(\bar{\gamma
})\right)  \label{qwall}%
\end{equation}
and so the ratio of safety factor at the wall to that at the axis is%
\begin{equation}
\frac{q_{wall}}{q_{axis}}=\frac{J_{1}(\bar{\gamma})}{J_{2}(\bar{\gamma})}%
\frac{\cos^{-1}\left(  J_{0}(\bar{\gamma})\right)  }{\pi}
\label{qwall by qaxis}%
\end{equation}
which is plotted in Fig.1. Contrary to Ref.\cite{Bellan2002} it is seen that
the shear (difference between $q_{wall}$ and $q_{axis})$ decreases with
increasing $\beta_{rel}$ (i.e., with $\bar{\gamma}$ decreasing below 2.405).
Using $i\theta=\ln\left(  \cos\theta+i\sin\theta\right)  $ to write
\begin{equation}
\cos^{-1}\left(  J_{0}(\bar{\gamma})\right)  =-i\ln\left(  J_{0}(\bar{\gamma
})+i\sqrt{1-\left(  J_{0}(\bar{\gamma})\right)  ^{2}}\right)
\label{eval inverse cosine}%
\end{equation}
and then using $J_{0}(\bar{\gamma})=1-\bar{\gamma}^{2}/4$ for $\bar{\gamma}%
\ll1$, it is seen that for $\bar{\gamma}\ll1$%
\begin{equation}
\cos^{-1}\left(  J_{0}(\bar{\gamma})\right)  \simeq-i\ln\left(  1-\frac
{\bar{\gamma}^{2}}{4}+i\frac{\bar{\gamma}}{\sqrt{2}}\right)  \simeq\frac
{\bar{\gamma}}{\sqrt{2}}. \label{eval inverse cosine 2}%
\end{equation}
Since $J_{1}(\bar{\gamma})\simeq\bar{\gamma}/2$ and $J_{2}(\bar{\gamma}%
)\simeq\bar{\gamma}^{2}/8$ for $\bar{\gamma}\ll1,$ Eq.\ref{qwall by qaxis} has
the limiting behavior
\begin{equation}
\ \frac{q_{wall}}{q_{axis}}\rightarrow\frac{4}{\sqrt{2}\pi}=0.900\text{ for
}\bar{\gamma}\ll1 \label{qwall by qaxis small gamma}%
\end{equation}
which is seen in Fig. 1. Furthermore, Eq.\ref{beta spheromak} has the limiting
behavior%
\begin{equation}
\ \beta_{rel}\rightarrow\frac{8}{\bar{\gamma}^{2}}\text{ for }\bar{\gamma}%
\ll1; \label{brel small gamma}%
\end{equation}
i.e., $\beta_{rel}$ diverges at small $\bar{\gamma}\ $\ which is also seen in
Fig. 1.

We note that numerical calculations reported in Ref.\cite{Gautier1981} assumed
$I^{2}\sim\psi^{2}(1+2\alpha\psi/3)$ and $dP/d\psi\sim\psi-\psi_{0}$ in a
spherical geometry and found that the gradient of the shear had a strong
dependence on $\alpha.$ The analytic solution given here would correspond
approximately to the $\alpha=0$ numerical solution reported in
Ref.\cite{Gautier1981}; the correspondence is not exact because of the
different assumptions for the dependence of $P$ on $\psi$, the shape of the
boundary (cylinder v. sphere), and the assumption of a central hole in Ref.
\cite{Gautier1981}.

\section{Toroidal magnetic bubble: Negative $\beta_{rel}$ \label{bubble}}

We now consider the situation where $\beta_{rel}<0$ and $\bar{\psi}_{lc}%
\neq0.$ We consider the $\sigma_{Y}=0$ case first as was assumed for
spheromaks and then later consider the more general case where both
$\sigma_{J}$ and $\sigma_{Y}$ are finite.

\subsection{$\sigma_{Y}=0$ case \label{sigmaY = 0}}

In the $\sigma_{Y}=0$ case $\bar{\psi}(\bar{r},\bar{z})$ is mathematically
identical to the spheromak solution considered in Sec.IV, i.e.,
Eq.\ref{solve GS alt} provides the relevant flux function. The difference here
is that $\bar{\psi}_{lc}$ is no longer assumed to be zero. Plots of $\bar
{\psi}(\bar{r},\bar{z})$ using $\bar{\gamma}>2.405$ show that $\bar{\psi}%
(\bar{r},0)$ has periodic maxima and minima because of its $J_{1}(\bar{\gamma
}\bar{r})$ dependence. Equation \ref{norm} defined $\bar{\psi}$ to be unity on
the magnetic axis, i.e., $\bar{\psi}(\bar{r},0)=1$ at $\bar{r}=1$ and the
magnetic axis was defined to be where $\bar{\psi}$ was a maximum or minimum.
Because of the oscillatory behavior of Bessel functions, maxima or minima of
$\bar{\psi}\,\ $occur not only at $r=1$ but also for $\bar{r}>1.$ However, the
maxima and minima occurring where $\bar{r}>1$ do not have $\bar{\psi}=1$ and
so do not satisfy the $\bar{\psi}\ =1$ condition given in Eq.\ref{norm}. Thus,
only the maximum of $\bar{\psi}(\bar{r},0)$ at $\bar{r}=1$ will be considered
since maxima or minima at larger $\bar{r}$ do not satisfy the $\bar{\psi}\ =1$
requirement stipulated in Eq.\ref{norm}.

Examination of Eq.\ref{solve GS alt} shows that $\bar{\psi}$ is independent of
$\bar{z}$ if $J_{1}(\bar{\gamma}\bar{r})=0;$ at this radius $\bar{r}%
=x_{11}/\bar{\gamma}$ where $x_{11}=3.832$ is the first root of $J_{1}.$ We
now show that this radius $\bar{r}=x_{11}/\bar{\gamma}$ is infinitesimally
larger than the radius of the last closed flux surface. Since $\bar{\psi}$ is
independent of $\bar{z}$ $\ $ when $J_{1}(\bar{\gamma}\bar{r})=0,$ the flux
surface passing through $\bar{r}=x_{11}/\bar{\gamma},$ $\bar{z}=0$ must be a
straight vertical line, i.e., $\bar{\psi}(x_{11}/\bar{\gamma},\bar{z}%
)=\bar{\psi}(x_{11}/\bar{\gamma},0)$ for all $\bar{z}.$ Because a straight
vertical line goes to $\bar{z}=\pm\infty$, the flux surface passing through
$\bar{r}=x_{11}/\bar{\gamma},\bar{z}=0$ is open. Immediately to the left of
this line the flux surfaces are closed and so the last closed flux surface is
at the radius $\bar{r}_{lc}$ where%
\begin{equation}
\bar{r}_{lc}=\lim_{\delta\rightarrow0}\left(  \frac{x_{11}}{\bar{\gamma}%
}-\delta\right)  =\frac{x_{11}}{\bar{\gamma}}. \label{rlc}%
\end{equation}
This can also be seen graphically from the flux surface contours shown in
Fig.2 (to be discussed in more detail later) where it is seen that a straight
vertical line separatrix lies between the blue-purple closed flux surfaces
having magnetic axis at $\bar{r}=1,$ $\bar{z}=0$ and the green-orange flux
surfaces to the right. Equation \ref{rlc} gives the radial location of this
vertical line.

A toroidal inverse aspect ratio (ratio of torus minor to major radius)\ can be
defined as%
\begin{equation}
\varepsilon=\frac{r_{lc}-r_{axis}}{r_{axis}}=\bar{r}_{lc}-1=\frac{x_{11}%
-\bar{\gamma}}{\bar{\gamma}}. \label{aspect}%
\end{equation}

Using $J_{1}(\bar{\gamma}\bar{r}_{lc})=0$ at the last closed flux surface, Eq.
\ref{solve GS alt} may be evaluated at $\bar{r}=\bar{r}_{lc},\bar{z}=0$ to
give
\begin{equation}
\bar{\psi}_{lc}=-\frac{x_{11}^{2}}{\bar{\gamma}^{2}}\frac{J_{0}(\bar{\gamma}%
)}{J_{2}(\bar{\gamma})}. \label{psi lc}%
\end{equation}
Inserting $\bar{\psi}_{lc}$ in Eq.\ref{beta spheromak} gives
\begin{equation}
\beta_{rel}=\left(  1+\frac{x_{11}^{2}}{\bar{\gamma}^{2}}\frac{J_{0}%
(\bar{\gamma})}{J_{2}(\bar{\gamma})}\right)  \frac{J_{0}(\bar{\gamma})}%
{J_{2}(\bar{\gamma})}. \label{neg beta}%
\end{equation}
In order to have $r_{lc}>r_{axis}$ Eq.\ref{aspect} shows that it is necessary
to have\ $\bar{\gamma}<x_{11}=3.832.$ A plot of Eq.\ref{neg beta} shows that
$\beta_{rel}$ is negative if $2.405<\bar{\gamma}<3.736$; $\beta_{rel}$ changes
sign at $\bar{\gamma}=3.736$ because the quantity in parenthesis in
Eq.\ref{neg beta} changes sign at $\bar{\gamma}=3.736.$ Thus if $2.405<\bar
{\gamma}<3.736$, $\beta_{rel}$ is negative and also \ $r_{lc}>r_{axis}.\ $

Because the minimum of $(1+s)s$ occurs when $s=-1/2,$ identifying
\ $s=x_{11}^{2}J_{0}(\bar{\gamma})/\left(  \bar{\gamma}^{2}J_{2}(\bar{\gamma
})\right)  $ it is seen that $\beta_{rel}$ is at a minimum when $x_{11}%
^{2}J_{0}(\bar{\gamma})/\left(  \bar{\gamma}^{2}J_{2}(\bar{\gamma})\right)
=-1/2$ in which case
\begin{equation}
\min\left[  \beta_{rel}\right]  =-\frac{\bar{\gamma}^{2}}{4x_{11}^{2}}.
\label{betamin}%
\end{equation}
$\ $

\bigskip Using the Bessel identities, the magnetic field components are$\ $
\begin{subequations}
\label{Bcomp unnorm J only}%
\begin{align}
B_{r}  &  =-\frac{\psi_{axis}}{2\pi r_{axis}^{2}\bar{r}}\frac{\partial
\bar{\psi}}{\partial\bar{z}}=\frac{\psi_{axis}}{2\pi r_{axis}^{2}\ }%
\frac{2\bar{k}\ }{\bar{\gamma}}\frac{J_{1}(\bar{\gamma}\bar{r})}{J_{2}%
(\bar{\gamma})}\sin(\bar{k}\bar{z})\label{Br solve}\\
B_{\phi}  &  =\frac{\lambda\psi}{2\pi r}=\ \frac{\psi_{axis}}{2\pi
r_{axis}^{2}\ }\ \frac{\bar{\lambda}}{J_{2}(\bar{\gamma})}\left(  \frac
{2}{\bar{\gamma}}J_{1}(\bar{\gamma}\bar{r})\cos(\bar{k}\bar{z})-\bar{r}%
J_{0}(\bar{\gamma})\right) \label{Bphi solve}\\
B_{z}  &  =\frac{\psi_{axis}}{2\pi r_{axis}^{2}\bar{r}}\frac{\partial\bar
{\psi}}{\partial\bar{r}}=\frac{\psi_{axis}}{2\pi r_{axis}^{2}}\ \frac{2}%
{J_{2}(\bar{\gamma})}\left(  \ J_{0}(\bar{\gamma}\bar{r})\cos(\bar{k}\bar
{z})-J_{0}(\bar{\gamma})\right)  . \label{Bz solve}%
\end{align}
Using Eq.\ref{identity} and Eq.\ref{Bphi solve} it is seen that $\ $%
\end{subequations}
\begin{equation}
B_{\phi,axis}=\frac{\psi_{axis}}{2\pi r_{axis}^{2}\ }\ \bar{\lambda}.
\label{Bphi axis.}%
\end{equation}

A normalized magnetic field can be defined as $\mathbf{\bar{B}=B(}\bar{r}%
,\bar{z}\mathbf{)/}B_{\phi,axis}$ with components
\begin{subequations}
\label{Bcomp norm J only}%
\begin{align}
\bar{B}_{r}(\bar{r},\bar{z})  &  =\frac{2\bar{k}\ }{\bar{\gamma}\bar{\lambda}%
}\frac{J_{1}(\bar{\gamma}\bar{r})}{J_{2}(\bar{\gamma})}\sin(\bar{k}\bar
{z})\label{Br norm}\\
\bar{B}_{\phi}(\bar{r},\bar{z})  &  =\frac{\frac{2}{\bar{\gamma}}J_{1}%
(\bar{\gamma}\bar{r})\cos(\bar{k}\bar{z})-\bar{r}J_{0}(\bar{\gamma})}%
{J_{2}(\bar{\gamma})\ }\label{Bphi norm}\\
\bar{B}_{z}(\bar{r},\bar{z})  &  =\frac{2}{\bar{\lambda}\ }\frac{\left(
J_{0}(\bar{\gamma}\bar{r})\cos(\bar{k}\bar{z})-J_{0}(\bar{\gamma})\right)
}{J_{2}(\bar{\gamma})}. \label{Bz norm}%
\end{align}
As required, both $\bar{B}_{r}$ and $\bar{B}_{z}$ vanish on the magnetic axis
(i.e., at $\bar{r}=1,\bar{z}=0$) and $\bar{B}_{\phi}=1$ on the magnetic axis.

Equation \ref{solve GS alt} with $2.405<\bar{\gamma}<3.736$ thus gives the
flux surface for a magnetic bubble, i.e., a toroidal configuration with closed
field lines where the pressure on the magnetic axis is \textit{lower} than the
pressure at the surface of the toroid. The direction of the $\mathbf{J\times
B}$ force is thus outwards rather than inwards in contrast to a tokamak. This
configuration is relevant to axisymmetric interplanetary magnetic clouds
ejected from the sun by coronal mass ejections. Spacecraft measurements
indicate that $P$ is smaller in the interior of these clouds than outside
$\ $so these clouds have negative $\beta_{rel}.$ Another possible situation
would be in the solar interior where a toroidal bubble configuration as
described here would be a toroidal region of stronger magnetic field but
reduced hydrodynamic pressure compared to the surroundings.

As a concrete example of such a configuration, consider the situation where
$\bar{\gamma}=\bar{k}=2.5$ and $\bar{\lambda}=\sqrt{2}\bar{\gamma}.$ In this
case $e=1$ so the poloidal flux surfaces are circular near the magnetic axis,
the last closed flux surface is at $\bar{\psi}_{lc}=0.25$ and from
Eq.\ref{neg beta} $\beta_{rel}=-0.081.$ From Eq.\ref{aspect}, it is seen that
the inverse aspect ratio is $\varepsilon=\allowbreak0.53.$ Figure 2 plots
contours of $\psi(\bar{r},\bar{z})$ and it is seen that the last closed flux
surface intersects $\bar{z}=0$ to the right of the magnetic axis at indeed
$\bar{r}_{lc}=x_{11}/\bar{\gamma}=\allowbreak1.\,\allowbreak53.$ Figures 3, 4,
5, and 6 plot $\bar{\psi}(\bar{r},0),$ $\bar{B}_{\phi}(\bar{r},0),$ $\bar
{B}_{z}(\bar{r},0),$ and $\bar{B}^{2}(\bar{r},0)\ $respectively.

From Eq.\ref{beta} it is seen that
\end{subequations}
\begin{equation}
\frac{\mu_{0}}{B_{axis}^{2}}P_{axis}=\frac{\mu_{0}}{B_{axis}^{2}}P_{lc}%
+\beta_{rel} \label{axis}%
\end{equation}
so the hydrodynamic pressure on the magnetic axis is lower than on the last
closed flux surface. If $P_{axis}$ is set to zero, then the external pressure
would be
\begin{equation}
\frac{\mu_{0}}{B_{axis}^{2}}P_{lc}=-\beta_{rel} \label{Plc}%
\end{equation}
in which case the configuration would be a vacuum at the magnetic axis (zero
plasma pressure) with increasing pressure going away from the magnetic axis
toward last closed flux surface.

If $\bar{\gamma}$ is further increased, the sign of $\beta_{rel}$ can become
positive again in which case the equilibrium will become tokamak-like (higher
pressure on magnetic axis). Additional increase of $\bar{\gamma}$ will cause
$\beta_{rel}$ to oscillate in sign giving a sequence of bubble-like and
tokamak-like configurations. Also, for a given configuration one could elect
to truncate the flux at some value larger than $\psi_{lc}$ and so obtain a
smaller aspect ratio equilibrium. In accordance with the Shafranov virial
theorem, any one of these configurations will involve a jump in the magnetic
field at the surface of the toroid if it is assumed that at the surface the
external magnetic field differs from the internal field. This jump corresponds
to the existence of surface currents. In a tokamak these surface currents are
provided by a set of coils immediately external to the toroidal volume and
these coils are called the vertical field coils. The field produced by these
coils is mainly in the $z$ direction and will be referred to here as
$B_{z}^{ext}$. This field $B_{z}^{ext}$ constitutes a portion of the total
field inside the toroidal volume and provides equilibrium in the major radius
direction. This takes place via a radial force $\sim J_{\phi}B_{z}^{ext}$
directed towards $\bar{r}=0\ $that balances the radially outward hoop force as
well as some hydrodynamic pressure forces. The hoop force is a property of any
toroidal current system and occurs because a toroidal current produces a
stronger poloidal field near $\bar{r}=0$, $\bar{z}=0$ (inside)\ than at
$\bar{r}=\bar{r}_{lc},$ $\bar{z}=0\ $(outside)$.$ This stronger poloidal field
on the inside compared to the outside corresponds to greater magnetic pressure
on the inside than on the outside; for low $\beta$ the force resulting from
magnetic pressure imbalance dominates any hydrodynamic pressure imbalance.
Without the offsetting force provided by $B_{z}^{ext}$ the hoop force would
act to expand the torus major radius.

At first sight it might appear that the flux contours in Fig.2 are such that
the magnetic pressure is higher on the outside than on the inside because the
midplane poloidal flux surfaces in Fig. 2 are more tightly packed outside the
magnetic axis (e.g., at $\bar{r}\simeq1.5)$ than inside the magnetic axis
(e.g., at $\bar{r}\simeq0.25$). However, the density of field lines and hence
the poloidal field is nevertheless stronger inside the magnetic axis than
outside because of the inverse $\bar{r}$ dependence in $B_{z}=(2\pi
r)^{-1}\partial\psi/\partial r$. The twice as tight midplane flux surface
packing in Fig. 2 at $\bar{r}\simeq1.5\ $ compared to at $\bar{r}\simeq0.25$
gives a twice as large $\partial\psi/\partial r$ on the outside compared to
the inside. However, this twice as tight radial packing is overcome by the
$(2\pi r)^{-1}$ factor, a toroidal geometry effect that produces an
approximately six-fold inside-to-outside enhancement with the net result that
$\left\vert B_{z}\right\vert $ is about three times larger at $\bar{r}%
\simeq0.25$ than at $\bar{r}\simeq1.5$. This three-fold inside-to-outside
ratio of $\left\vert B_{z}\right\vert $ is evident in Fig. 5. $\ $

In order to have the $B_{z}^{ext}$ required \ for equilibrium, it would be
necessary to have surface currents flowing on the surface of the toroid. Since
there are no powered coils to sustain surface currents exterior to a magnetic
cloud, it is unlikely that such surface currents would exist in the magnetic
cloud context. Without the $B_{z}^{ext}$ provided by surface currents \ (and
intrinsic to the equilibrium given here), the hoop force resulting from the
imbalance between $B_{z}^{2}$ on the inside and $B_{z}^{2}$ on the outside
will cause the major radius of magnetic clouds to increase with time. The
difference between poloidal flux surfaces with and without incorporation of
$B_{z}^{ext}$ is of the order of the inverse aspect ratio $\varepsilon$
because $B_{z}^{ext}\ $ is a toroidal effect and so scales as $\varepsilon.$

\subsection{Finite $\sigma_{J}$ and $\sigma_{Y}$ case}

The spheromak solution required $\sigma_{Y}$ to be zero to avoid singularity
at $\bar{r}=0.$ The magnetic bubble solution discussed above used the same
functional form as the spheromak solution (i.e., had $\sigma_{Y}=0$ and used
Eq.\ref{GS solve}) and found that a tokamak-like solution with $\beta_{rel}<0$
$\ $(i.e., inverted beta profile) occurred if $2.405<\bar{\gamma}<3.736.$ If
$\bar{r}=0$ is excluded from the domain so the configuration is
\textit{doubly-connected}, the singular nature of $Y_{1}(\bar{\gamma}\bar{r})$
at $\bar{r}=0$ is no longer a constraint and the more general solution given
by Eqs. \ref{GS soln}, \ref{solve sigmaJ}, and \ref{solve sigmaY} can be used.
The consequence of imposing $\sigma_{Y}=0$ was for Eq.\ref{solve sigmaY} to
force the relationship \ between $Q$ and $\bar{\gamma}$ given by Eq.\ref{Q}.
If $\sigma_{Y}$ is not forced to be zero, then this relationship between $Q$
and $\bar{\gamma}$ is no longer imposed and the only remaining condition is
that the domain must exclude $\bar{r}=0.$

Consideration of Eq.\ref{GS soln} and recalling the discussion that led to
Eq.\ref{rlc} shows that $\bar{\psi}(\bar{r}_{lc},\bar{z})$ is independent of
$\bar{z}$ at $\bar{r}_{lc}$ where $\bar{r}_{lc}$ is now defined by
\begin{equation}
\sigma_{J}J_{1}(\bar{\gamma}\bar{r}_{lc})+\sigma_{Y}Y_{1}(\bar{\gamma}\bar
{r}_{lc})=0. \label{rlc more general}%
\end{equation}
Thus Eq.\ref{rlc more general} provides a radial shift of the location of the
last closed flux surface and generalizes the discussion that led to
Eq.\ref{rlc}. Because $\sigma_{J}$ and $\sigma_{Y}$ depend on $\bar{\gamma}$
and on $Q$ (hence on $\beta_{rel}$), Eq.\ref{rlc more general} shows that
$\bar{r}_{lc}$ depends on both $\bar{\gamma}$ and $\beta_{rel}.$ However, by
assumption $\bar{r}_{lc}>1$ (last closed flux surface radius is to the right
of the magnetic axis) which restricts the allowed values of $\bar{\gamma}$ and
$\beta_{rel}$. Introduction of the $Y_{1}(\bar{\gamma}\bar{r})$ solution and
the coefficients $\sigma_{J}$ and $\sigma_{Y}$ is thus analogous to
generalizing the solution of some harmonic equation from being $\sin(kx)$ to
being $\sin(kx+\Delta)=\sin(kx)\cos(k\Delta)+\cos(kx)\sin(k\Delta)$ where
$\sin(kx)$, $\cos(kx)$ are the analogs of $J_{1}(\bar{\gamma}\bar{r}%
),Y_{1}(\bar{\gamma}\bar{r})$ and $\sigma_{J},\sigma_{Y}$ are the analogs of
$\cos(k\Delta),\sin(k\Delta).$ Introducing finite $\Delta$ changes the phase
of the solution and shifts the location of the solution.

Substitution for $\sigma_{J}$ and $\sigma_{Y}$ in Eq.\ref{rlc more general}
using Eqs.\ref{solve sigmaJ}, \ref{solve sigmaY} gives%
\begin{equation}
Q(\bar{\gamma},\bar{r}_{lc})=\frac{Y_{0}(\bar{\gamma})J_{1}(\bar{\gamma}%
\bar{r}_{lc})-J_{0}(\bar{\gamma})Y_{1}(\bar{\gamma}\bar{r}_{lc})}{Y_{2}\left(
\bar{\gamma}\right)  J_{1}(\bar{\gamma}\bar{r}_{lc})\ -J_{2}(\bar{\gamma
})Y_{1}(\bar{\gamma}\bar{r}_{lc})}. \label{solve Q most general}%
\end{equation}
It is seen that Eq.\ref{solve Q most general} reduces to Eq.\ref{Q} if
$J_{1}(\bar{\gamma}\bar{r}_{lc})=0$, i.e., the situation considered in
Sec.\ref{sigmaY = 0} and that $Q$ becomes infinite when $\bar{r}_{lc}$ is
\ such that the denominator in the right hand side of
Eq.\ref{solve Q most general} vanishes.

Using Eqs.\ref{rlc more general} in Eq.\ref{GS soln} it is seen that the last
closed flux surface is given by
\begin{equation}
\bar{\psi}_{lc}=-\bar{r}_{lc}^{2}Q \label{solve lc gen}%
\end{equation}
and inserting this in Eq.\ref{def Q} gives%
\begin{equation}
\beta_{rel}=Q+\bar{r}_{lc}^{2}Q^{2}. \label{betarel Q}%
\end{equation}
The derivative of Eq.\ref{betarel Q} shows that the minimum possible
$\beta_{rel}$ is $\beta_{rel}=-1/(4\bar{r}_{lc}^{2})$ which occurs when
$Q=-1/(2\bar{r}_{lc}^{2});$ this generalizes \ Eq.\ref{betamin}.

Figure 7 plots the dependence of $Q$ and $\beta_{rel}$ on $\bar{r}_{lc}$ for
$1<\bar{r}_{lc}<2$ with $\bar{\gamma}=2.5;$ it is seen that, as predicted,
$\beta_{rel}$ has a minimum at $\beta_{rel}=-1/(4\bar{r}_{lc}^{2})$ which
occurs when $Q=-1/(2\bar{r}_{lc}^{2}).$ It is also seen from this figure that
when $\bar{r}_{lc}=x_{11}/\bar{\gamma}=1.5328$ the Sec.\ref{sigmaY = 0} result
$\beta_{rel}=-0.081$ and $Q=J_{0}(\bar{\gamma})/J_{2}(\bar{\gamma})=$
$-0.108\ $ is recovered. For this $\bar{\gamma}=2.5$ value the denominator in
Eq.\ref{solve Q most general} vanishes when $\bar{r}_{lc}\rightarrow1.885.$

The following chain of dependence thus exists for doubly-connected configurations:\ 

\begin{enumerate}
\item Independent values for $\bar{r}_{lc}$ and $\bar{\gamma}$ can be selected
which then determine $Q$ via Eq. \ref{solve Q most general},

\item Using Eqs.\ref{solve sigmaJ} and \ref{solve sigmaY} in Eq.\ref{GS soln}
the flux function \ $\bar{\psi}(\bar{r},\bar{z})$ is given by
\begin{equation}
\bar{\psi}(\bar{r},\bar{z})=\frac{\pi}{2}\bar{\gamma}\bar{r}\left\{
\begin{array}
[c]{c}%
\,\left[  Y_{0}(\bar{\gamma})-QY_{2}\left(  \bar{\gamma}\right)  \right]
J_{1}(\bar{\gamma}\bar{r})\\
+\left[  -J_{0}(\bar{\gamma})+QJ_{2}(\bar{\gamma})\right]  Y_{1}(\bar{\gamma
}\bar{r})
\end{array}
\right\}  \cos(\bar{k}\bar{z})-\bar{r}^{2}Q, \label{psi general}%
\end{equation}

\item $\beta_{rel}$ is given by Eq.\ref{betarel Q},

\item $\bar{\psi}_{lc}$ is given by Eq.\ref{solve lc gen}.
\end{enumerate}

This chain of dependence for doubly-connected configurations differs from that
of a finite $\beta$ spheromak. Specifically the chain of dependence for a
finite $\beta$ spheromak is: $\bar{\psi}_{lc}=0$ is imposed because of the
singly-connected topology, $\bar{r}_{lc}$ is determined from setting the left
hand side of Eq.\ref{solve GS alt}$\ $to zero on the midplane, and
Eq.\ref{beta spheromak} gives $\beta_{rel}=J_{0}(\bar{\gamma})/J_{2}%
(\bar{\gamma})$.

Another and equivalent point of view differentiating singly- and
doubly-connected configurations from each other is the following:

(i) because the midplane of a singly-connected configuration contains $\bar
{r}=0$ and because $\bar{\psi}=0$ at $\bar{r}=0$, the last closed flux surface
for a singly-connected\ configuration must always have $\bar{\psi}=0,$

whereas in contrast,

(ii) because the midplane of a doubly-connected configuration excludes
$\bar{r}=0,$ the last closed flux surface of a doubly-connected configuration
cannot be $\bar{\psi}=0$ as such a flux surface would have to pass through
$\bar{r}=0.$

The magnetic field components associated with Eq.\ref{psi general} normalized
to $B_{\phi,axis}=\bar{\lambda}\psi_{axis}/\left(  2\pi r_{axis}^{2}\right)  $
are $\ $
\begin{subequations}
\label{Bcomp norm J and Y}%
\begin{align}
\bar{B}_{r}(\bar{r},\bar{z})  &  =\frac{\bar{k}}{\bar{\lambda}}\frac{\pi
\bar{\gamma}}{2}\left\{
\begin{array}
[c]{c}%
\,\left[  Y_{0}(\bar{\gamma})-QY_{2}\left(  \bar{\gamma}\right)  \right]
J_{1}(\bar{\gamma}\bar{r})\\
+\,\left[  -J_{0}(\bar{\gamma})+QJ_{2}(\bar{\gamma})\right]  Y_{1}(\bar
{\gamma}\bar{r})
\end{array}
\right\}  \sin(\bar{k}\bar{z})\label{Br norm JY}\\
\bar{B}_{\phi}(\bar{r},\bar{z})  &  =\frac{\pi\bar{\gamma}}{2}\left\{
\begin{array}
[c]{c}%
\,\left[  Y_{0}(\bar{\gamma})-QY_{2}\left(  \bar{\gamma}\right)  \right]
J_{1}(\bar{\gamma}\bar{r})\\
+\,\left[  -J_{0}(\bar{\gamma})+QJ_{2}(\bar{\gamma})\right]  Y_{1}(\bar
{\gamma}\bar{r})
\end{array}
\right\}  \cos(\bar{k}\bar{z})-\bar{r}Q\label{Bphi norm JY}\\
\bar{B}_{z}(\bar{r},\bar{z})  &  =\frac{1}{\bar{\lambda}\ }\left(  \frac
{\pi\bar{\gamma}^{2}}{2}\left\{
\begin{array}
[c]{c}%
\,\left[  Y_{0}(\bar{\gamma})-QY_{2}\left(  \bar{\gamma}\right)  \right]
J_{0}(\bar{\gamma}\bar{r})\\
+\,\left[  -J_{0}(\bar{\gamma})+QJ_{2}(\bar{\gamma})\right]  Y_{0}(\bar
{\gamma}\bar{r})
\end{array}
\right\}  \cos(\bar{k}\bar{z})-2Q\right)  . \label{Bz norm JY}%
\end{align}
Using Eqs.\ref{identity} and \ref{Wronskian} it is seen that
Eq.\ref{Bcomp norm J and Y} reverts to Eq.\ref{Bcomp norm J only} when
$Q=J_{0}(\bar{\gamma})/J_{2}(\bar{\gamma}).$

\textit{Acknowledgements:} This material is based upon work supported by the
U.S. Department of Energy Office of Science, Office of Fusion Energy Sciences
under Award Numbers DE-FG02-04ER54755 and DE-SC0010471, by the National
Science Foundation under Award Number 1059519, and by the Air Force Office of
Scientific Research under Award Number FA9550-11-1-0184. The authors wish to
thank an anonymous reviewer for making the valuable suggestion that $Y_{n}$
Bessel solutions be allowed when $\bar{r}=0$ is excluded from the domain.

\bigskip

\pagebreak
\end{subequations}

\pagebreak

\bigskip

\bigskip

\bigskip\
%TCIMACRO{\FRAME{ftbpFU}{3.659in}{2.1292in}{0pt}{\Qcb{$q_{wall}/q_{axis}$
%plotted as black solid line v. $\bar{\gamma}$ from Eq.\ref{qwall by qaxis} and
%$\beta_{rel}$ plotted as red dashed line.}}{\Qlb{qaxis by qwall}%
%}{pop_fig1_new3.eps}{\special{ language "Scientific Word";  type "GRAPHIC";
%width 3.659in;  height 2.1292in;  depth 0pt;  original-width 3.6115in;
%original-height 2.0903in;  cropleft "0";  croptop "1";  cropright "1";
%cropbottom "0";  filename 'PoP_fig1_new3.eps';file-properties "XNPEU";}}}%
%BeginExpansion
\begin{figure}[ptb]%
\centering
\includegraphics[
height=2.1292in,
width=3.659in
]%
{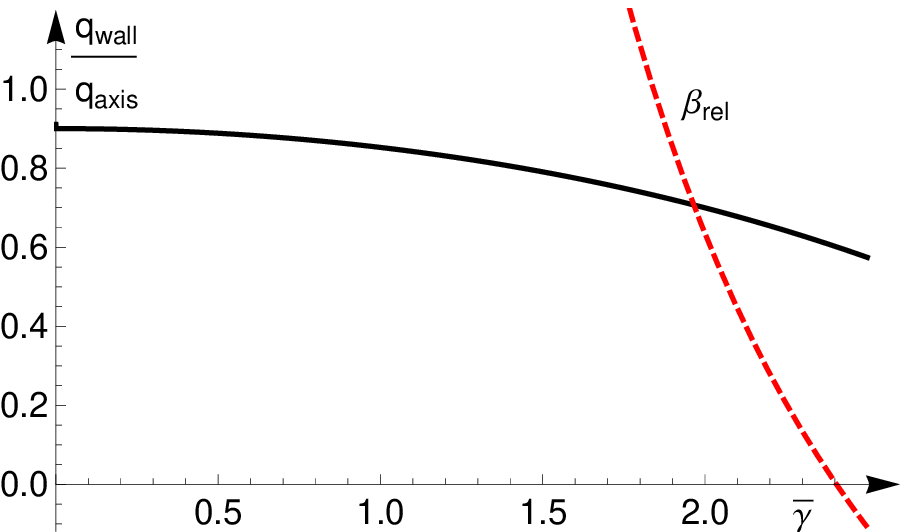}%
\caption{$q_{wall}/q_{axis}$ plotted as black solid line v. $\bar{\gamma}$
from Eq.\ref{qwall by qaxis} and $\beta_{rel}$ plotted as red dashed line.}%
\label{qaxis by qwall}%
\end{figure}
%EndExpansion%
%TCIMACRO{\FRAME{ftbpFU}{3.659in}{3.5025in}{0pt}{\Qcb{Contour plot of
%$\bar{\psi}(\bar{r},\bar{z})$ given by Eq.\ref{solve GS alt} for $\bar{\gamma
%}=\bar{k}=2.5.$ The radius of the last closed flux surface is at $\bar
%{r}=1.53.$ The hydrodynamic pressure is lower in the pink region than in the
%green region so the configuration is a magnetic bubble.}}{\Qlb{contour}%
%}{pop_fig2_hue.eps}{\special{ language "Scientific Word";  type "GRAPHIC";
%width 3.659in;  height 3.5025in;  depth 0pt;  original-width 3.6115in;
%original-height 3.4558in;  cropleft "0";  croptop "1";  cropright "1";
%cropbottom "0";  filename 'PoP_fig2_hue.eps';file-properties "XNPEU";}}}%
%BeginExpansion
\begin{figure}[ptb]%
\centering
\includegraphics[
height=3.5025in,
width=3.659in
]%
{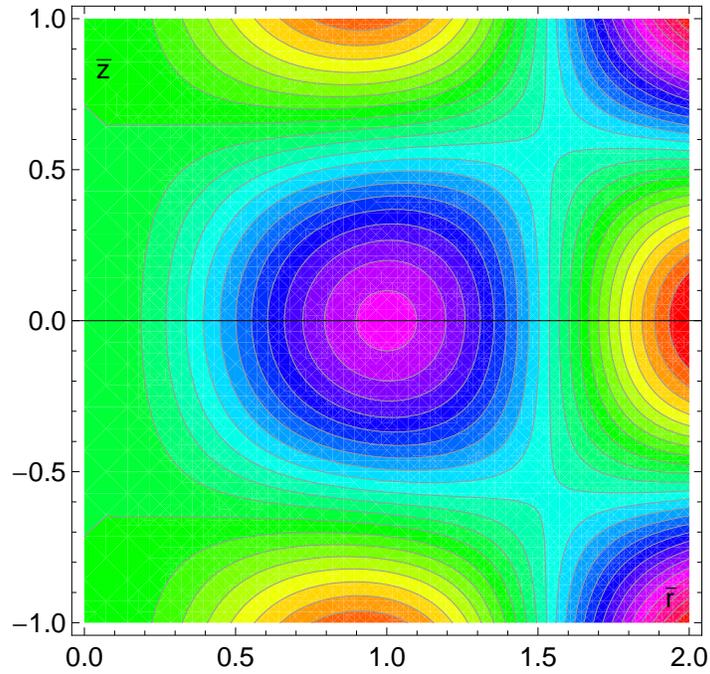}%
\caption{Contour plot of $\bar{\psi}(\bar{r},\bar{z})$ given by
Eq.\ref{solve GS alt} for $\bar{\gamma}=\bar{k}=2.5.$ The radius of the last
closed flux surface is at $\bar{r}=1.53.$ The hydrodynamic pressure is lower
in the pink region than in the green region so the configuration is a magnetic
bubble.}%
\label{contour}%
\end{figure}
%EndExpansion%
%TCIMACRO{\FRAME{ftbpFU}{3.659in}{2.0894in}{0pt}{\Qcb{Plot of $\bar{\psi}%
%(\bar{r},0)$ as given by Eq.\ref{solve GS alt} for $\bar{\gamma}=\bar{k}=2.5$
%and $\bar{\lambda}=2\sqrt{\bar{\gamma}}.$ The radius of the last closed flux
%surface is at $\bar{r}=1.53\ $and $\bar{\psi}_{lc}=0.25.$}}{\Qlb{psi midplane}%
%}{pop_fig3_new3.eps}{\special{ language "Scientific Word";  type "GRAPHIC";
%width 3.659in;  height 2.0894in;  depth 0pt;  original-width 3.6115in;
%original-height 2.0505in;  cropleft "0";  croptop "1";  cropright "1";
%cropbottom "0";  filename 'PoP_fig3_new3.eps';file-properties "XNPEU";}}}%
%BeginExpansion
\begin{figure}[ptb]%
\centering
\includegraphics[
height=2.0894in,
width=3.659in
]%
{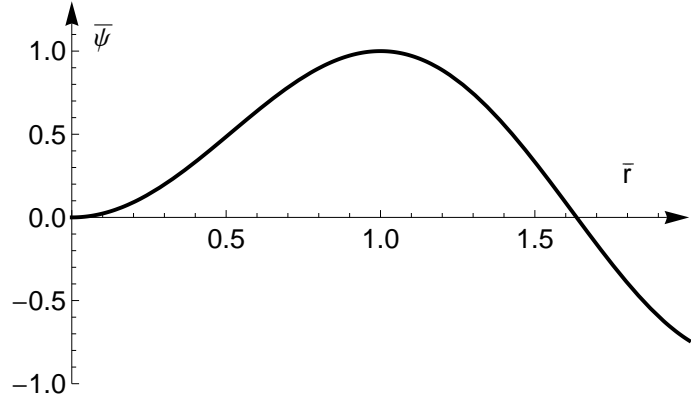}%
\caption{Plot of $\bar{\psi}(\bar{r},0)$ as given by Eq.\ref{solve GS alt} for
$\bar{\gamma}=\bar{k}=2.5$ and $\bar{\lambda}=2\sqrt{\bar{\gamma}}.$ The
radius of the last closed flux surface is at $\bar{r}=1.53\ $and $\bar{\psi
}_{lc}=0.25.$}%
\label{psi midplane}%
\end{figure}
%EndExpansion%
%TCIMACRO{\FRAME{ftbpFU}{3.659in}{2.3964in}{0pt}{\Qcb{$\bar{B}_{\phi}(\bar
%{r},0)$; Note that magnetic axis is at $\bar{r}=1.0$ and that maximum of
%$\bar{B}_{\phi}$ occurs to left of magnetic axis.}}{\Qlb{BPhi-on-axis}%
%}{pop_fig4_new.eps}{\special{ language "Scientific Word";  type "GRAPHIC";
%width 3.659in;  height 2.3964in;  depth 0pt;  original-width 3.6115in;
%original-height 2.3557in;  cropleft "0";  croptop "1";  cropright "1";
%cropbottom "0";  filename 'PoP_fig4_new.eps';file-properties "XNPEU";}}}%
%BeginExpansion
\begin{figure}[ptb]%
\centering
\includegraphics[
height=2.3964in,
width=3.659in
]%
{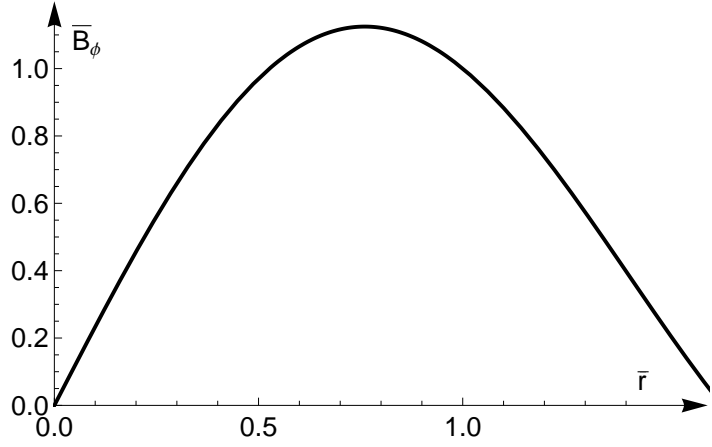}%
\caption{$\bar{B}_{\phi}(\bar{r},0)$; Note that magnetic axis is at $\bar
{r}=1.0$ and that maximum of $\bar{B}_{\phi}$ occurs to left of magnetic
axis.}%
\label{BPhi-on-axis}%
\end{figure}
%EndExpansion
%

%TCIMACRO{\FRAME{ftbpFU}{3.659in}{2.1439in}{0pt}{\Qcb{ $\bar{B}_{z}(\bar
%{r},0);$ note that $\bar{B}_{z}$ changes sign at magnetic axis.}%
%}{\Qlb{Bz-on-axis}}{pop_fig5_new.eps}{\special{ language "Scientific Word";
%type "GRAPHIC";  maintain-aspect-ratio TRUE;  display "USEDEF";
%valid_file "F";  width 3.659in;  height 2.1439in;  depth 0pt;
%original-width 3.6115in;  original-height 2.1049in;  cropleft "0";
%filename 'PoP_fig5_new.eps';file-properties "XNPEU";}}}%
%BeginExpansion
\begin{figure}[ptb]%
\centering
\includegraphics[
height=2.1439in,
width=3.659in
]%
{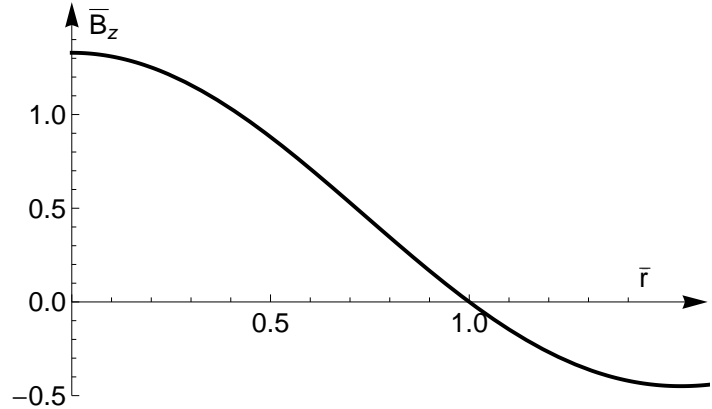}%
\caption{ $\bar{B}_{z}(\bar{r},0);$ note that $\bar{B}_{z}$ changes sign at
magnetic axis.}%
\label{Bz-on-axis}%
\end{figure}
%EndExpansion%
%TCIMACRO{\FRAME{ftbpFU}{3.659in}{2.3964in}{0pt}{\Qcb{$\bar{B}_{r}^{2}+\bar
%{B}_{\phi}^{2}+\bar{B}_{z}^{2}$ as function of $\bar{r}$ for $\bar{z}=0.$}%
%}{\Qlb{Bsquared-on-axis}}{pop_fig6_new3.eps}%
%width 3.659in;  height 2.3964in;  depth 0pt;  original-width 3.6115in;
%original-height 2.3557in;  cropleft "0";  croptop "1";  cropright "1";
%cropbottom "0";  filename 'PoP_fig6_new3.eps';file-properties "XNPEU";}}}%
%BeginExpansion
\begin{figure}[ptb]%
\centering
\includegraphics[
height=2.3964in,
width=3.659in
]%
{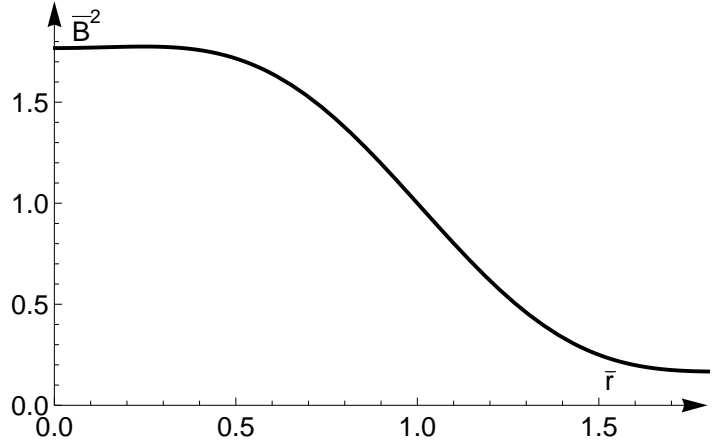}%
\caption{$\bar{B}_{r}^{2}+\bar{B}_{\phi}^{2}+\bar{B}_{z}^{2}$ as function of
$\bar{r}$ for $\bar{z}=0.$}%
\label{Bsquared-on-axis}%
\end{figure}
%EndExpansion%
%TCIMACRO{\FRAME{ftbpFU}{5.0416in}{2.9477in}{0pt}{\Qcb{Figure 7. Blacksolid
%line is plot of $Q(\gamma,\bar{r}_{lc})$ as function of $\bar{r}_{lc}$ with
%$\bar{\gamma}=2.5;$ red dashed line is plot of $\beta_{rel}.$ For this
%$\bar{\gamma},$ $Q$ diverges when $\bar{r}_{lc}=1.885.$}}{\Qlb{Qplot}%
%}{pop_fig7.eps}{\special{ language "Scientific Word";  type "GRAPHIC";
%width 5.0416in;  height 2.9477in;  depth 0pt;  original-width 5.3761in;
%original-height 3.1309in;  cropleft "0";  croptop "1";  cropright "1";
%cropbottom "0";  filename 'PoP_fig7.eps';file-properties "XNPEU";}}}%
%BeginExpansion
\begin{figure}[ptb]%
\centering
\includegraphics[
height=2.9477in,
width=5.0416in
]%
{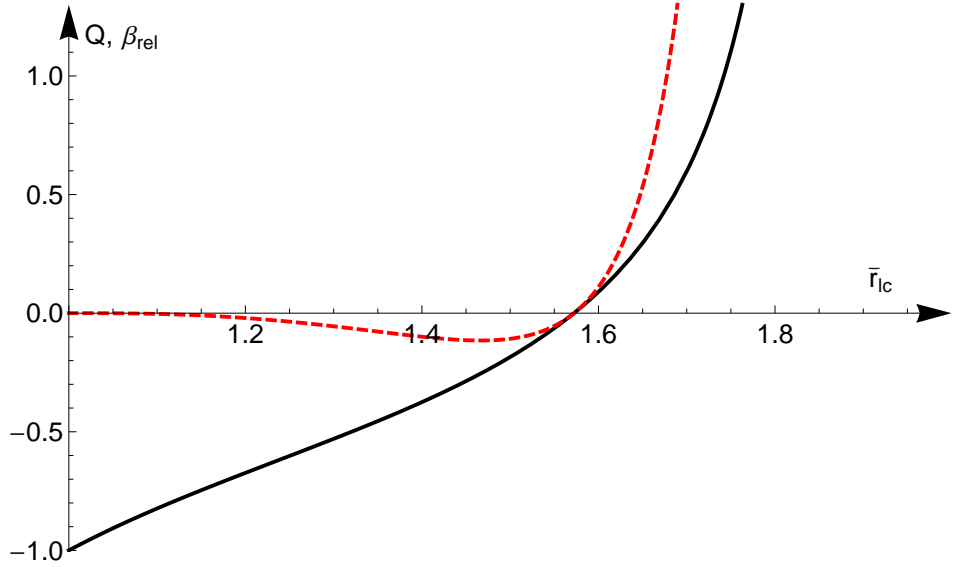}%
\caption{Figure 7. Blacksolid line is plot of $Q(\gamma,\bar{r}_{lc})$ as
function of $\bar{r}_{lc}$ with $\bar{\gamma}=2.5;$ red dashed line is plot of
$\beta_{rel}.$ For this $\bar{\gamma},$ $Q$ diverges when $\bar{r}%
_{lc}=1.885.$}%
\label{Qplot}%
\end{figure}
%EndExpansion

\pagebreak
\end{document}